\documentclass[12pt,letterpaper]{article}
\usepackage[utf8x]{inputenc}
\usepackage{ucs}
\usepackage{amsmath}
\usepackage{amsfonts}
\usepackage{amssymb}
\usepackage{graphicx}
\usepackage[left=2cm,right=2cm,top=2cm,bottom=2cm]{geometry}
\usepackage[pdftex,bookmarks=true,colorlinks=true,citecolor=blue,plainpages=false]{hyperref}
\usepackage[round]{natbib}

\author{Angelo Mele
	\footnote{This version: \today . I am grateful to Zhenling Jiang, Cornelius Fritz and Philip Solimine for helpful discussions. All remaining errors are mine.}
}

\title{Estimating Network Models using Neural Networks\\
EXTENDED ABSTRACT}

\makeindex 
\begin{document}

\maketitle

\begin{abstract}
Exponential random graph models (ERGMs) are very flexible for modeling network formation but pose difficult estimation challenges due to their intractable normalizing constant. Existing methods, such as MCMC-MLE, rely on sequential simulation at every optimization step. We propose a neural network approach that trains on a single, large set of parameter–simulation pairs to learn the mapping from parameters to average network statistics. Once trained, this map can be inverted, yielding a fast and parallelizable estimation method. The procedure also accommodates extra network statistics to mitigate model misspecification. Some simple illustrative examples show that the method performs well in practice. 
\end{abstract}

\section{Introduction}

Exponential random graphs (ERGMs) provide a flexible modeling strategy for networks, by assuming that the probability of observing a network is proportional to an exponential function of network statistics (linear or nonlinear), potentially incorporating features such as edges, triangles, mutual links, or exogenous covariates \citep{Snijders2002, robins2007introduction}. 
In the linear (canonical) case, this defines a standard exponential family distribution for the network; in the non-linear case, we have a curved exponential family \citep{Mele2017}.

Recent work in the econometrics of networks has shown that ERGMs have a microeconomic foundation as equilibrium of stochastic best-response dynamics. \citep{Mele2017} shows that the ERGM is the stationary equilibrium of such best-response dynamics, thus allowing a structural interpretation of the parameters as payoffs of a well-defined potential game of network formation \citep{MondererShapley2006}. The linear (or nonlinear) function in the exponential distribution thus summarizes the incentives of the players to form, maintain or delete links.\footnote{Other structural models in economics follow similar ideas but do not exploit a potential game characterization, so these models are not ERGMs \citep{DePaula2017, DePaulaEtAl2018, Graham2017} }

While the flexibility of the specfication makes the ERGM framework extremely popular in applications and among practitioners, the estimation of this model is quite challenging. The main bottleneck is that the likelihood of the model is proportional to an intractable normalizing constant, which sums the exponential function over all possible network configurations, a number that is exponential in the number of possible links. The usual workaround involves Monte Carlo-based methods (MCMC-MLE) to approximate the likelihood via network simulations. These simulations can be computationally expensive, especially for larger networks or complex dependence terms \citep{BhamidiEtAl2011, ChatterjeeDiaconis2013}. In principle, parallelization can help simulate multiple chains, but the MCMC-MLE optimization process is typically iterative and sequential, thus restricting how far parallelism can reduce the total runtime.

An alternative view is to target the network statistics in a method-of-moments or minimum-distance framework, but such methods also require repeated simulation at different parameter guesses. Similarly to MCMC-MLE, the optimization is sequential, so one has to wait the optimization step before running another set of simulated networks, thus reducing potential benefits of parallel computer architecture. 

In this paper, I propose an estimation approach that leverages simple neural networks (NN) to approximate the map from parameters to expected network statistics off-line in a single, parallelizable training step \citep{JiangWei2025, goodfellow2016deep}. Once the neural network is trained, we can invert the map (via a suitable optimization) to find the parameter that best matches the observed network's statistics. By separating the simulation stage (where parallelization is straightforward) from the inference stage, this approach can significantly reduces the sequential bottleneck common to MCMC-based methods.

Additionally, the map generated by the neural network can be used to study identification and discover areas of the parameter space that lead to degeneracy, a common problem of this class of models. 

The rest of the paper is organized as follows. 
Section~\ref{sec:ergm} introduces the ERGM framework and highlights examples ranging from simple Erd\H{o}s-R\'enyi to more complex terms like mutual edges, triangles, or covariates.
Section~\ref{sec:estimation} reviews standard ERGM estimation techniques, including maximum pseudolikelihood, MCMC-MLE, and contrastive divergence. 
Section~\ref{sec:nn-approach} describes the estimation procedure in detail, explaining how to train a network to learn the mapping from parameters to moments and then invert it. 
Finally, Section~\ref{sec:conclusion} summarizes potential advantages, limitations, and future directions.

\section{Exponential Random Graph Models}
\label{sec:ergm}

Exponential random graphs assume that the probability of observing network $g$ with parameter vector $\theta$ is given by
\begin{equation}
    \pi(g; \theta) \;=\;
    \frac{\exp\bigl(Q(g;\theta)\bigr)}{c(\theta)},
\end{equation}
where $Q(g;\theta)$ is a potential function that depends on network statistics and parameters, and the normalizing constant $c(\theta)$ is
\begin{equation}
    c(\theta) \;=\; \sum_{\omega\in\mathcal{G}} \exp\bigl(Q(\omega;\theta)\bigr),
\end{equation}
with $\mathcal{G}$ the set of all possible networks on $n$ nodes. Directly computing $c(\theta)$ is infeasible for all but very small $n$ because $|\mathcal{G}| = 2^{\binom{n}{2}}$ for an undirected network (or $2^{n(n-1)}$ for directed).

When $Q(g;\theta)$ is \emph{linear} in $\theta$ and in the network statistics, $\pi(g;\theta)$ is a canonical exponential family. Nonlinear or \emph{curved} terms lead to more complicated sufficiency structures but remain within the ERGM class.

%

In practice, ERGMs can combine multiple network statistics, each weighted by a corresponding parameter. Here we outline some of the most common terms:

\begin{enumerate}
    \item \textbf{Erd\H{o}s--R\'enyi (Bernoulli) Model.}

    This is arguably the simplest ERGM, in which every edge is independent of the others and has the same probability of forming. The sufficient statistic is the number of links
    \[
    t_e(g) \;=\; \text{(total number of edges in }g\text{).}
    \]
    The probability of observing a network $g$ takes the form
    \[
    \pi(g;\theta) 
    \;=\;
    \frac{\exp\bigl(\theta \, t_e(g)\bigr)}{\displaystyle\sum_{\omega \in \mathcal{G}} 
    \exp\bigl(\theta \, t_e(\omega)\bigr)},
    \]
    where $\theta \in \mathbb{R}$ is the (scalar) parameter controlling edge density. A positive $\theta$ implies a higher expected number of edges (i.e., more connected network), whereas a negative $\theta$ leads to a sparser network.

    \item \textbf{Mutual Edges (Directed Networks).}

    In a directed network, we may care not only about how many edges exist but also about how often edges are \emph{reciprocated}. We thus define the number of mutual links
    \[
    t_{m}(g) 
    \;=\; 
    \sum_{i < j} g_{ij}g_{ji},
    \]
    i.e., the number of \emph{mutual dyads}. Here $g_{ij}$ indicates whether node $i$ has an outgoing edge to node $j$. In an ERGM with edges and reciprocity, we have a potential 
    \[
    Q(g;\theta) 
    \;=\; \theta_1 \, t_e(g) 
         \;+\; \theta_2 \, t_{m}(g),
    \]
    reflecting a linear combination of total directed edges and mutual edges. A positive $\theta_2$ encourages reciprocal relationships.

    \item \textbf{Triangles, Transitivity, and GWESP.}

    \begin{itemize}
      \item \emph{Triangles:} In an undirected graph, define 
      \[
      t_{\triangle}(g) 
      \;=\;
      \sum_{i < j < k} 
        g_{ij} g_{jk}g_{ik},
      \]
      i.e., the total count of “triangles” (3-cliques). Including $t_{\triangle}(g)$ in the potential function 
      \[
      Q(g;\theta)
      \;=\; \theta_1 t_e(g) + \theta_2 t_{\triangle}(g)
      \]
      helps capture transitive closure or clustering in a network. Here a positive $\theta_2$ favors more clustering. 

      \item \emph{Geometrically Weighted Edgewise Shared Partners (GWESP):    CHECK THIS FOR THE FORMULA} 
      A common curved ERGM term for transitivity is GWESP (geometrically weighted edgewise shared partners). If $g_{i,j}$ is an edge, let $s_{ij}$ be the number of shared partners that $i$ and $j$ have. Then the GWESP statistic can be written as
      \[
      t_{\mathrm{GWESP}}(g;\alpha)
      \;=\;
      \sum_{(i,j) \in E(g)} 
      \Bigl(1 - (1 - \alpha)^{\, s_{ij}}\Bigr),
      \]
      where $\alpha$ is a decay parameter ($0 < \alpha < 1$). A positive coefficient on GWESP means the model favors configurations where edges share more neighbors (transitivity), but the geometric decay $\alpha$ penalizes high counts of shared partners in a specific way.
    \end{itemize}

    \item \textbf{Covariates (Node-Level or Dyadic).}

    We can incorporate exogenous characteristics at the node level or at the edge level:
    \begin{enumerate}
      \item \emph{Node-Level Covariates:} Suppose each node $i$ has an attribute $x_i$ (e.g.\ gender, income, etc.). A simple homophily statistic could be:
      \[
      t_{x}(g)
      \;=\;
      \sum_{i < j} \mathbf{1}\bigl(x_i = x_j\bigr)\, g_{ij},
      \]
      which counts edges between nodes sharing the same attribute.
      \item \emph{Dyadic Covariates:} If each pair $(i,j)$ has a covariate $z_{ij}$, we might set
      \[
      t_{z}(g)
      \;=\;
      \sum_{i < j} z_{ij} g_{ij}.
      \]
      Positive coefficient implies the model favors edges where $z_{ij}$ is larger.
    \end{enumerate}
    More elaborate forms are possible, e.g.\ interactions with structural terms or non-linear transformations of covariates.
\end{enumerate}

In general, each sufficient statistic $t_k(g)$ captures a structural feature of the network. In our examples we have: 
\begin{itemize}
  \item \emph{Edges} measure overall density; 
  \item \emph{Mutual dyads} capture reciprocity in directed graphs; 
  \item \emph{Triangles / transitivity} capture clustering (friends-of-friends also becoming friends); 
  \item \emph{Covariates} measure external or attribute-driven tendencies to form ties.
\end{itemize}

In an ERGM, the model coefficient $\theta_k$ for each statistic $t_k(\cdot)$ indicates whether the presence (or higher value) of that feature is encouraged ($\theta_k>0$) or discouraged ($\theta_k<0$), holding other statistics fixed.

\section{Existing Methods of Estimation}
\label{sec:estimation}
In this section, I provide a quick and non-exhaustive review of the existing methods for ERGM estimation. The main point is that all these methods require \emph{sequential} likelihood or moment evaluations, each of which often uses MCMC. Although parallelization using multiple chains is possible, the optimizer itself must still wait for each iteration, limiting speed. 

\subsection{Maximum Pseudolikelihood Estimation (MPLE)}

A classical issue in ERGMs is the intractable normalizing constant, which appears in the full likelihood. One shortcut is to exploit the conditional probability that an edge $(ij)$ is present, given the rest of the network. Specifically, define the conditional probability that $i$ and $j$ have a link, conditioning on the rest of the network as
\[
P\bigl(g_{ij}=1 \,\mid\, g_{-ij}; \theta\bigr),
\]
where $g_{-ij}$ denotes all edges and configurations except $g_{ij}$. In many ERGMs -- especially the ones with linear potential function -- this conditional probability can be written in a form similar to a logistic regression,
\[
\texttt{logit}\Bigl(
P(g_{ij}=1 \mid g_{-ij};\theta)
\Bigr) 
\;=\; 
\theta^\top \Delta_{ij}(g_{-ij}),
\]
where $\Delta_{ij}(g_{-ij})$ is a vector of differences in sufficient statistics when $g_{ij}$ changes from 0 to 1, holding all other edges fixed. 

The \emph{pseudolikelihood} is constructed by assuming these conditional probabilities factorize as if the presence/absence of each edge were independent. This yields a product form,
\[
\mathcal{L}_{\text{PL}}(\theta) 
\;=\;
\prod_{i<j} 
P\bigl(g_{i,j}=1 \mid g_{-(i,j)}; \theta\bigr)^{g_{i,j}}
\,
\bigl[1 - P(g_{i,j}=1 \mid g_{-(i,j)}; \theta)\bigr]^{\,1-g_{i,j}}.
\]
Maximizing $\log \mathcal{L}_{\text{PL}}(\theta)$ with respect to $\theta$ is akin to running a large logistic regression with response $g_{i,j}$ and covariates $\Delta_{i,j}(g_{-(i,j)})$. Hence, MPLE is straightforward and fast to compute using standard GLM routines available in most statistical software packages.

The advantage of MPLE is evident, as it avoids directly dealing with the normalizing constant $c(\theta)$. It can be fit in many software packages as a logistic regression problem, making it computationally appealing for moderately sized networks. Once the $\Delta_{i,j}$ statistics are constructed, standard routines can estimate $\theta$ quickly.

However, by factorizing over edges, the pseudolikelihood treats each edge event as conditionally independent given the rest of the network. This ignores higher-order dependencies that may exist (e.g. transitivity, reciprocity). This can lead to severe bias in models with strong structural effects.

While MPLE is consistent in certain limited scenarios (e.g.\ purely dyadic independence), for general ERGMs with dependence terms, the theoretical guarantees are unclear or do not hold \citep{BoucherMourifie2017}. Consequently, there is no universal result ensuring the consistency of MPLE for large networks, especially if the model includes complex terms like triangles or GWESP.

Finally, it is well known that even in simpler models, MPLE estimates can be systematically off from the true maximum likelihood estimates. Standard errors derived from naive logistic regression output do not account for the mismatch between pseudolikelihood and the true likelihood. Specialized variance corrections or bootstrap approaches may be required to get more honest confidence intervals.

Despite these limitations, MPLE remains widely used because of its speed and ease of implementation. It often serves as a good starting point for MCMC-MLE or for exploratory analysis, especially in high-dimensional settings where full MCMC-based methods are costly. However, for final inference in models with strong dependence terms (like transitivity), one must be aware of potential bias and the need for adjustments or alternative estimators.

\subsection{MCMC Maximum Likelihood (MCMC-MLE)}
The classical likelihood-based method uses Markov chain Monte Carlo to approximate the gradient or log-likelihood ratio. Given a candidate $\theta$, one simulates networks from $\pi(\cdot;\theta)$ and compares them to the observed network. An iterative algorithm (e.g. Robbins--Monro, stochastic approximation, gradient ascent) refines $\theta$ to a new guess $\theta^{\prime}$ and a new simulation is started from $\theta^{\prime}$; and so on. 
Although consistent, MCMC-MLE can be computationally expensive because each iteration requires fresh simulations, and some parameter regions can lead to degenerate simulated networks -- either all complete or empty networks. In such cases the likelihood approximation becomes unstable and the optimization may fail.

In MCMC-MLE, at a high level, one seeks to solve
\[
\hat{\theta}
\;=\;
\arg \max_{\theta}
\;\log \pi\bigl(g_{\text{obs}};\theta\bigr),
\]
where
\[
\log \pi\bigl(g_{\text{obs}};\theta\bigr)
\;=\;
Q\bigl(g_{\text{obs}};\theta\bigr)
-\log c(\theta).
\]
Since $c(\theta)$ is the sum over all possible networks, a Markov chain Monte Carlo (MCMC) approximation is typically used to estimate gradients or likelihood ratios.

The typical MCMC-MLE algorithm proceeds according to the following steps:
\begin{enumerate}
    \item \textbf{Initialize} with some parameter guess $\theta^{(0)}$.
    \item \textbf{At iteration $k$:} 
    \begin{enumerate}
        \item Simulate $R$ networks from $\pi(\cdot;\theta^{(k)})$ using a suitable MCMC sampler (e.g., Gibbs or Metropolis--Hastings).
        \item Approximate the gradient $\nabla \log \pi(g_{\text{obs}};\theta^{(k)})$ by comparing the observed network statistics to the average statistics of the simulated networks.
        \item Update $\theta^{(k+1)} = \theta^{(k)} + \gamma_k \cdot \widehat{\nabla}$ 
        (a Robbins--Monro or stochastic gradient approach).
    \end{enumerate}
    \item \textbf{Repeat} until convergence, e.g., until $\theta^{(k)}$ changes negligibly or a maximum number of iterations is reached.
\end{enumerate}
Over many iterations, the algorithm refines $\theta$ to maximize the approximate likelihood.

    For sufficiently large $R$ (number of MCMC samples) and a well-behaved ERGM (no extreme degeneracy), MCMC-MLE converges to the true maximum likelihood estimate \citep{GeyerThompson1992}. This is indeed the “gold standard” for inference when it can be done reliably.
    Because the method only needs to simulate from $\pi(\cdot;\theta)$, it can handle curved ERGMs, models with complicated statistics (GWESP, mutual, etc.), and large sets of covariates, provided that the Markov chain mixes reasonably well, avoiding degeneracy and getting trapped in local modes.

On the other hand, the algorithm presents some limitations. Indeed, each iteration requires generating new MCMC draws from the current parameter. If the chain needs a large burn-in or suffers slow mixing (common in heavily constrained or degenerate regions), running many iterations can be prohibitively expensive.
    Although one can parallelize \emph{within} each iteration (e.g., run multiple short MCMC chains to reduce variance), the overall algorithm is inherently sequential: one must wait for iteration $k$ before choosing $\theta^{(k+1)}$. This feature severely reduces how much parallelization can help speed-up the computations.
    Furthermore, for certain parameter values, the chain may produce almost all-empty or all-complete networks (near-degenerate samples). This can stall the algorithm, making gradient estimates noisy or misleading and requiring careful tuning of starting values or tempering schemes.

Finally, the Markov chain burn-in length, proposal distributions, and step size $\gamma_k$ in the Robbins--Monro update can significantly affect both speed and stability of convergence and require fine tuning. Suboptimal tuning can lead to slow or incorrect convergence.

Despite these limitations, MCMC-MLE is often viewed as the most principled approach to ERGM estimation, providing (under suitable conditions) consistent estimates and valid standard errors. For moderate-sized networks and non-degenerate parameter regions, it can perform well. However, the computational overhead and sequential nature motivate exploring alternative methods, such as the neural network approach proposed in this paper, which offloads most of the simulation to a single parallelizable stage.

\subsection{Bayesian Methods}
Another, more recent literature approaches the estimation problem from a Bayesian perspective. The idea is to use a double Metropolis-Hastings step to simulate from the posterior distribution of the ERGM parameters. This is accomplished using variants of the exchange algorithm \citep{MurrayEtAl2006, CaimoFriel2010, Liang2010, Mele2017}: when a new parameter is proposed, we simulate a network using a Metropolis-Hastings algorithm, and accept the parameter with high probability if the simulated network is similar to the observed one. The similarity is measured using likelihood ratios. This simulation scheme avoids computation of the constant in the posterior as well as in the likelihood, thus bypassing the bottleneck in estimation. However, such algorithms tend to mix very slowly, and convergence to the final posterior approximation crucially depends on the lenght of the network MCMC simulations. Since the posterior simulation is also sequential in nature, because MCMC are inherently sequential, this scheme may be computationally very costly.

\subsection{Other approaches}
Inspired by machine learning (e.g.\ training Boltzmann machines), CD approximates the needed gradient with relatively short MCMC chains, starting from the observed network (or other initial states). This is typically faster than full MCMC-MLE but introduces an additional approximation error dependent on chain length and initial conditions. 

On the other hand, \cite{MeleZhu2021} propose a mean-field approximation method, based on the analysis of \cite{ChatterjeeDiaconis2013}. The methods is computationally attractive, and while the authors show that the mean-field likelihood approximation error becomes vanishingly small with the size of the network, the statistical properties of the estimator are not fully understood, except in the case of simple models. 

\section{A Neural Network Approach to ERGM Estimation}
\label{sec:nn-approach}
The main limitation of the estimation approaches in the previous section is the sequential nature of the algorithms. 
The proposed approach in this paper is a two-stage procedure that first learns the mapping from  the vector of parameters $\theta$ to the vector of expected statistics of the network $E[t(g,\theta)]$, by training a Neural Network.
In the second stage, the algorithm finds $\widehat{\theta}$ by inverting this map. This is based on the fact that the MLE solves the following system of equations
\begin{eqnarray}
t(g^{obs}) = E[t(g,\theta)]
\end{eqnarray}
and one can find $\widehat{\theta}$ such that the equation is satisfied. Here the network statistics $t(g,\theta)$
 are the chosen sufficient statistics in the ERGM specification; however, the researcher can also add additional moments, to alleviate problems of misspecification or to improve the fit. A similar approach was recently suggested for structural models in the marketing literature \citep{JiangWei2025}.\\

In practice, the proposed algorithm works as follows:

\begin{enumerate}
    \item \textbf{Sampling Parameter Vectors.} 
    Choose a range or prior distribution over the parameter space $\Theta$, and draw $L$ parameter vectors $\{\theta^{(\ell)}\}_{\ell=1}^L$. 
    \item \textbf{Parallel Simulations.}
    For each $\theta^{(\ell)}$, simulate $M$ networks via standard MCMC methods \emph{independently and in parallel}. Compute the average statistics 
    \[
    \overline{t}\bigl(\theta^{(\ell)}\bigr) 
    \;=\; \frac{1}{M}\sum_{m=1}^M t\bigl(g^{(\ell,m)}\bigr).
    \]
    \item \textbf{Neural Network Training.}
    Collect the pairs $\bigl(\theta^{(\ell)}, \overline{t}(\theta^{(\ell)})\bigr)$ into a dataset. Train a neural network $f_\phi: \Theta \to \mathbb{R}^d$ that approximates 
    \[
    f_\phi\bigl(\theta^{(\ell)}\bigr) 
    \;\approx\; \overline{t}\bigl(\theta^{(\ell)}\bigr).
    \]
    \item \textbf{Parameter Estimation.}
    For the \emph{observed} network $g_\text{obs}$, compute its statistics $t_\text{obs} = t(g_\text{obs})$. Then find
    \[
    \hat{\theta} 
    \;=\; \arg \min_{\theta \in \Theta} \Bigl\| f_\phi(\theta) \,-\, t_\text{obs} \Bigr\|.
    \]
    This last step is a simple numerical optimization or “root-finding”, which can be done quickly since $f_\phi$ is just a feed-forward pass. This avoids repeated MCMC calls typical of the MCMC-MLE algorithm.
    
\end{enumerate}

There are several advantages to this approach. 
First, the $L$ training data with parameters and network statistics can each be simulated in parallel, fully exploiting multi-core or cluster computing. Therefore, the algorithm is now a embarassingly parallel problem.

Second, instead of repeatedly simulating networks at each iteration of an MCMC-MLE, the algorithm does a one-shot large design of parameter vectors, providing a more complete picture of the parameter space and corresponding network statistics.

Third, we can include additional network statistics to alleviate model misspecification or improve fit. This also allows the researcher to step away from the implicit assumption that the specification is correct.

One interesting feature is that while training the neural network we are also implicitly performing a goodness-of-fit check. This is useful because the goodness-of-fit is usually separated from the estimation. In practice the researcher usually estimates the model, then runs a long simulation to check the fit. If the fit is unsatisfactory, then one has to try different specifications or longer runs of the MCMC to obtain better fit. In the proposed algorithm there is no such step, as the fit is evaluated during the training step.

Finally, after training, finding $\hat{\theta}$ involves only forward passes through the neural network, which is fast even in high dimensions. Once the NN is trained, any model with the same specification can be estimated through this method, without need to run additional simulations.

\subsection{Illustrative Examples}

\begin{figure}[ht]
\caption{Predicted vs test data, model with edges only}
	\includegraphics[scale=0.2]{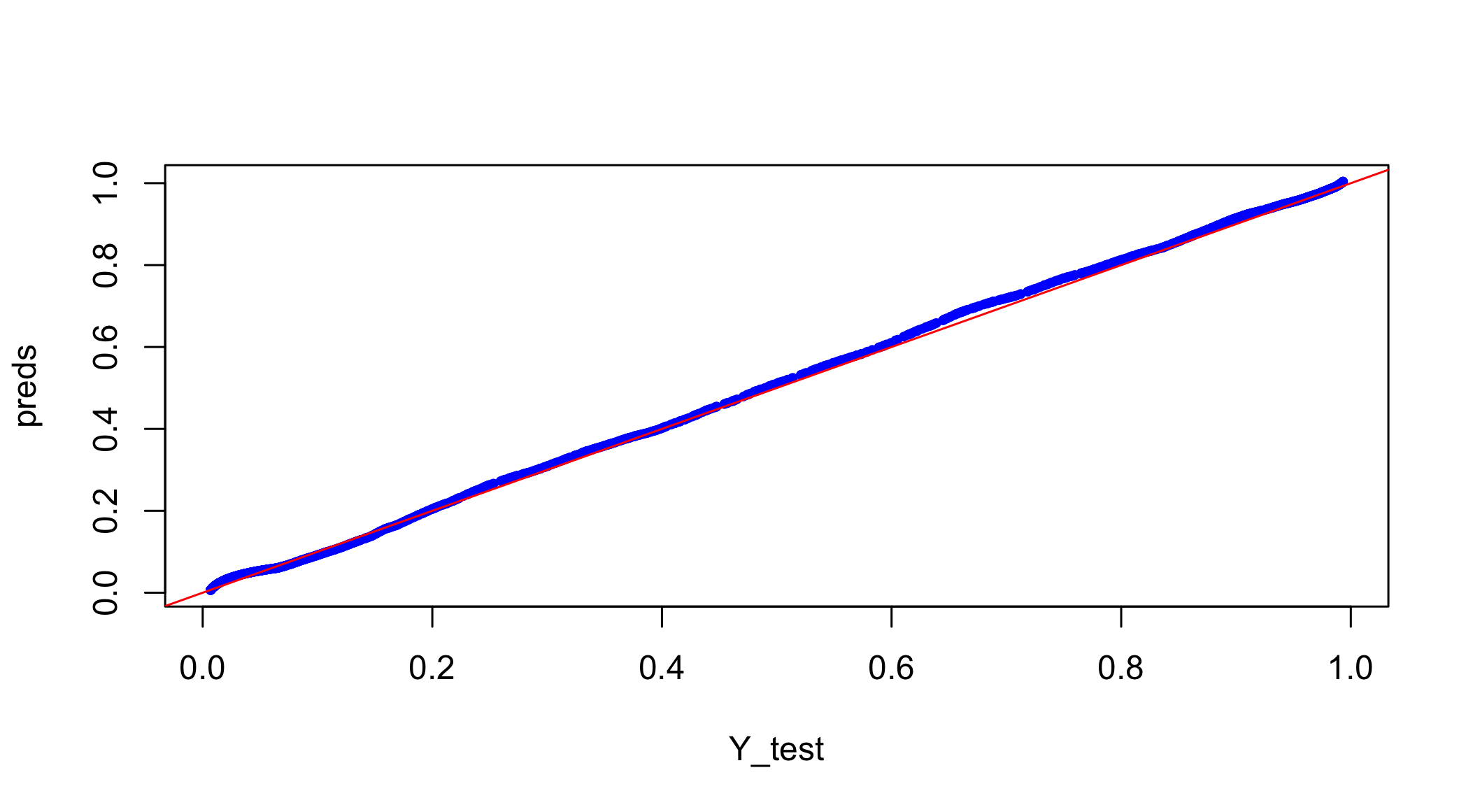}
	\label{fig:predicted_vs_test_edges_model}
\end{figure} 
We provide a few illustrative examples with some simple models. The starting model is an Erdos-Renyi graph, which has a single parameter. We generate simulated networks using parameters $\theta\in[-5,5]$. We sample $L=10,000$ parameter values from $U[-5,5]$ and for each parameter value we simulate $M=10,000$ networks, to compute the average number of edges (the sufficient statistics of the model). We then feed this dataset to a simple feedforward Neural Network with 
with three dense layers.\footnote{I use a very simple NN. The first hidden layer contains 128 neurons and uses the ReLU activation function, followed by a 20\% dropout layer to mitigate overfitting. The second hidden layer has 64 neurons with ReLU activation, again followed by a 20\% dropout layer. Finally, an output layer with a linear activation produces as many outputs as there are moments in the training data (in our case only one). I train the model for 200 epochs, and use 20\% of the data for validation testing.}

In Figure \ref{fig:predicted_vs_test_edges_model}, I report the performance of the NN in predicting the average number of links. In blue it is the predicted values and the red line represent the perfect fit. In general this simple model performs quite well.

\begin{figure}
	\caption{Training data vs. predicted vs Theoretical number of links, models with edges only}
	\includegraphics[scale = 0.2]{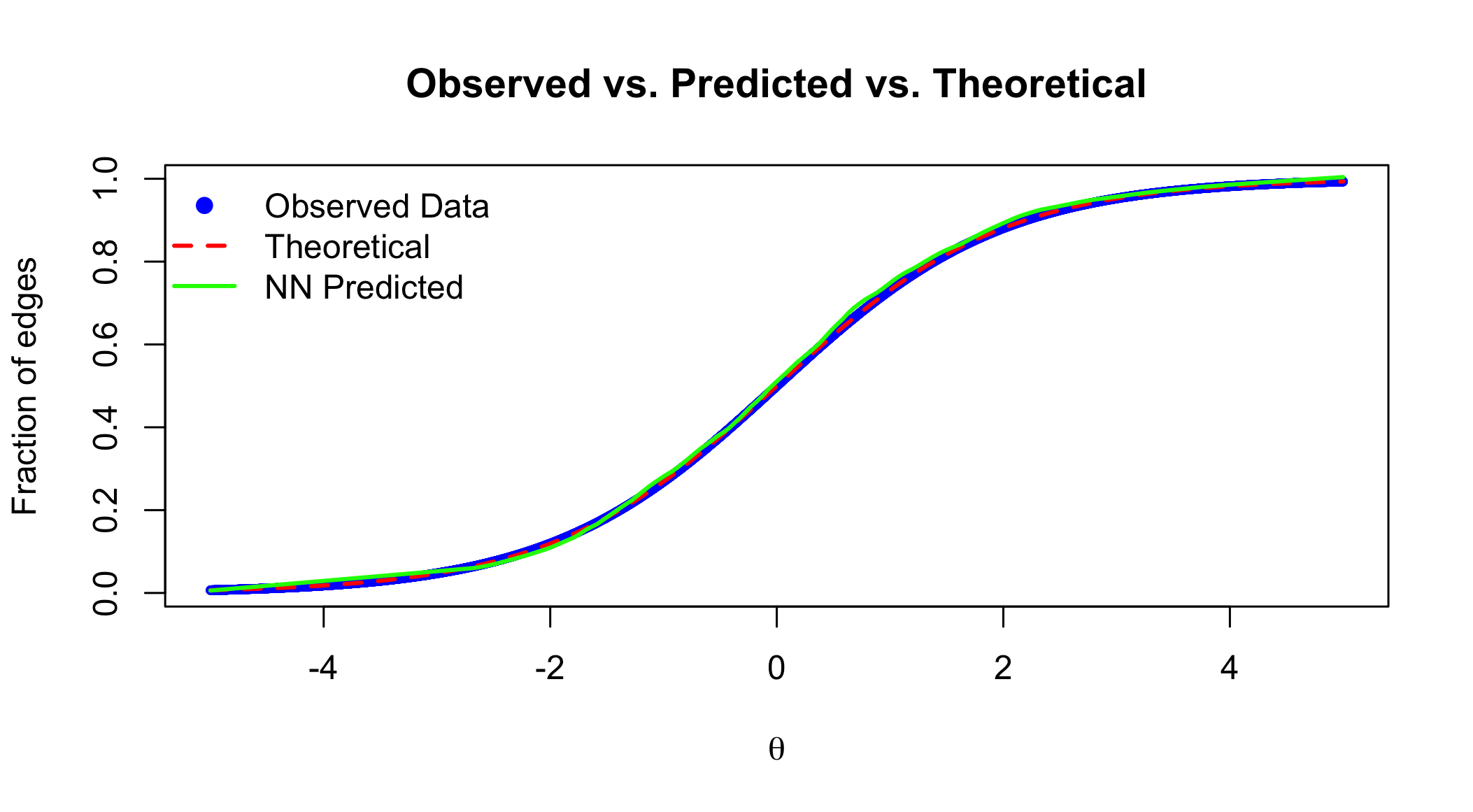}
	\label{fig:trainingdata_vs_predicted_vs_theoretical_edges_model}
\end{figure}

The same performance is observed in Figure \ref{fig:trainingdata_vs_predicted_vs_theoretical_edges_model}, where I plot the training data (blue) against the predicted values (red). Because this is an Erdos-Renyi graph we can also plot the theoretical value of the links (green). Again, the NN performs very well in this simple example. To obtain an estimate from the trained neural network, one has to just feed the model a value for the number of links and find the best parameter, which is an easy task according to the Figure.
    
\begin{figure}
\caption{Predicted vs test data, model with edges and mutual links}
	\includegraphics[scale=0.2]{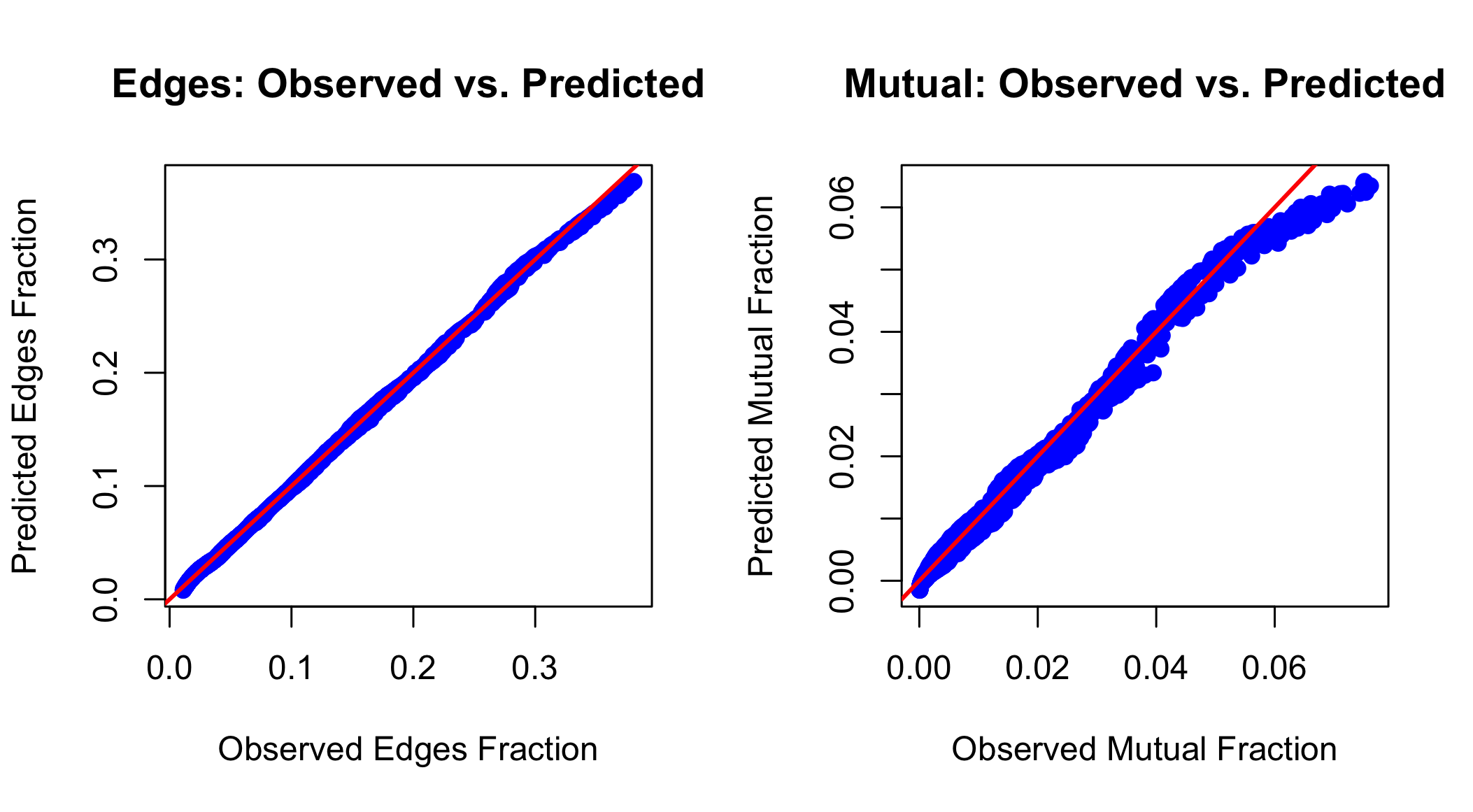}	\label{fig:predicted_vs_test_edges_mutual_model}
\end{figure}

We also test the procedure in the model with edges and mutual links, shown in Figure \ref{fig:predicted_vs_test_edges_mutual_model}.
Here we see that the prediction of the reciprocated ties is a little off for some parameter values. This suggest that we should probably increase the number of simulated samples to obtain a better training dataset.

In summary, the procedure performs relatively well for the simple models we considered in this proof-of-concept application. In terms of computation speed, the generation of the training data takes between 3 and 5 minutes on a cluster with 48 processors. The training takes a few seconds. 

\section{Discussion and Future Work}
\label{sec:conclusion}

This paper develops a neural network--based estimator (\textbf{NNERGM}) that shifts the burden of MCMC simulation into a one-time, parallelizable training dataset. By learning the map $\theta \mapsto E[t(\theta)]$ off-line, we avoid iterative simulations in the final inference step. This approach can provide a scalable alternative to MCMC-MLE for large $n$, if one has sufficient parallel compute resources. The methods adapt naturally to curved ERGMs or more complex dependence structures by including richer sets of statistics during training.
Furthermore, the neural step approach provides estimation and goodness-of-fit checks in a single step, while other algorithms separate the estimation from the goodness-of-fit.

Another interesting feature is that it can potentially reveal regions of parameter degeneracy if certain areas map to near-complete or near-empty networks, allowing users to refine the design or impose constraints. This would provide a way to check for identification of the models using the trained neural network.

A potential drawback is that if multiple distinct parameter vectors produce similar network statistics, the learned mapping may not be invertible in practice. However, this points to a modeling issue (identification) rather than a problem with the method. If the model is identifiable, then the method performs well and the map should be invertible.
 
Generating a large training set can still be expensive, although it is \emph{embarrassingly parallel}. A possible avenue to parsimonious simulations is to start with a training set around a focal vector of parameters, and then extend the simulations in areas that need more exploration or require more precision. While this heuristic will probably work, there is a need to provide theoretical guarantees and statistical properties to justify such simulation strategy.

Future work will include more examples and estimation with real data. A more compelling example would include unobserved heterogeneity at the node level (random or fixed effects).
To compute the standard errors, one can exploit the properties of the exponential family or run some additional simulation at the estimated parameter value. 

Overall, the NNERGM approach provides a promising new perspective on large-scale or complex ERGM estimation: \emph{simulate once, learn the map, and invert quickly.}

\bibliographystyle{jmr}	
\bibliography{references}		

\end{document}